\documentstyle[epsfig,12pt,twoside]{article}

\begin{document}

\begin{center}
{\bf\large Study of $N^*$(1440) from $J/\Psi$ Decays }

\vskip 1cm

{\bf B.S.Zou$^{a-d}$, G.X.Peng$^{a,b}$, R.G.Ping$^b$, H.C.Chiang$^{a-d}$, 
W.X.Ma$^{a,b}$ and P.N.Shen$^{a-d}$}

\bigskip
a) CCAST(World Lab.), P.O.Box 8730, Beijing 100080;\\
b) Institute of High Energy Physics, Chinese Academy of Science,\\ 
P.O.Box 918(4), Beijing 100039\footnote{Mailing address};\\
c) Institute of Theoretical Physics, Chinese Academy of Sciences\\
d) Center of Theoretical Nuclear Physics, National Laboratory of
Heavy Ion Accelerator, \\ Lanzhou 730000, P.R.China  \\
\end{center}

\begin{abstract}

For $J/\Psi\to\bar pp \pi^0$ and $\bar pp\pi^+\pi^-$, 
the $\pi^0p$ and $p\pi^+\pi^-$ systems 
are limited to be pure isospin 1/2 due to isospin conservation. This is
a big advantage in studying $N^*$ resonances from $J/\Psi$ decays,
compared with $\pi N$ and $\gamma N$ experiments. 
The process $J/\Psi\to\bar pN^*$ or $p\bar N^*$ provides a new way to
probe the internal structure of the $N^*$ resonances. 
Here we report a quark model calculation for
$J/\Psi\to\bar pp$, $\bar pN^*(1440)$ and $\bar N^*N^*$.
The implication for the internal structure of $N^*(1440)$ is discussed. 

\end{abstract}

\vspace{0.5cm}
\vspace{0.5cm}

\section{Introduction}
An important source of information about the nucleon internal structure
is the properties of nucleon excitation states $N^*$'s, such as their mass
spectrum, various production and decay rates\cite{Burkert1}. 
Our present knowledge of this aspect came almost entirely from
partial-wave 
analyses of $\pi N$ total, elastic, and charge-exchange scattering data of
more than twenty years ago\cite{PDG}.
Since the late 1970's, very little has happened in experimental $N^*$
baryon spectroscopy. Considering its importance for the understanding
of the baryon structure and for distinguishing various pictures
\cite{Isgur1} of the nonperturbative regime of QCD, a new generation
of experiments on $N^*$ physics with electromagnetic probes has recently
been started at new facilities such as CEBAF at JLAB, ELSA at Bonn,
GRAAL at Grenoble. 

A long-standing problem in $N^*$ physics is about the nature of the
Roper resonance $N^*(1440)$. In simple three-quark picture of baryons, it
should be the first radial excitation state of the nucleon. But various
quark models\cite{Isgur1} met difficulties to explain its mass and
electromagnetic couplings. It has therefore been suggested\cite{Hybrid} to
be a gluonic excitation state of the nucleon, i.e., a ``hybrid baryon".
To establish the gluonic degree of freedom in hadrons is a fascinating
challenge in nowadays non-perturbative QCD physics.

Although the existence of the $N^*(1440)$ is well-established, its 
properties, such as mass, width and decay branching ratios etc., still
suffer large experimental uncertainties\cite{PDG}.
A big problem in extracting information on the $N^*(1440)$ from $\pi N$
and $\gamma N$ experiments is the isospin decomposition of 1/2 and 3/2
\cite{Workman}. 
As pointed out by one of us\cite{Zou}, the decays of $J/\Psi\to \bar
pp\pi^0$ and $J/\Psi\to\bar pp\pi^+\pi^-$ provide an ideal place for 
studying the properties of $N^*$ resonances, since in these processes the
$\pi^0p$ and $p\pi^+\pi^-$ systems 
are limited to be pure isospin 1/2 due to isospin conservation. 
Preliminary results from the BES Collaboration on $J/\Psi\to\bar pp\pi^0$
show a clear peak structure around 1490 MeV in its $\pi^0p$ invariant
mass spectrum\cite{BES}.

The process $J/\Psi\to\bar pN^*$ or $p\bar N^*$ also provides a new way to
probe the internal quark-gluon structure of the $N^*$ resonances. 
In the simple 
three-quark picture of baryons, the process can be described by Fig.1
\cite{Carimalo}. In this picture, three quark-antiquark pairs are 
created independently via a symmetric three-gluon intermediate state
with no extra interaction other than the recombination process in the
final state to form baryons. This is quite different from the mechanism
underlying the $N^*$ production from the $\gamma p$ process where the
photon
couples to only one quark and unsymmetric configuration of quarks is
favored.   Therefore the processes $J/\Psi\to\bar pN^*$ and 
$\gamma p\to N^*$ should probe different aspects of the quark
distributions inside baryons. 
Since the $J/\Psi$ decay is a glue-rich process, a hybrid $N^*$ is 
expected to have larger production rate than a pure three-quark 
$N^*$\cite{Page}.

\vspace*{-2.cm}
\begin{figure}[htbp]
\vspace*{-2.cm}
\begin{center}\hspace*{-0.cm}
\epsfysize=12cm
\epsffile{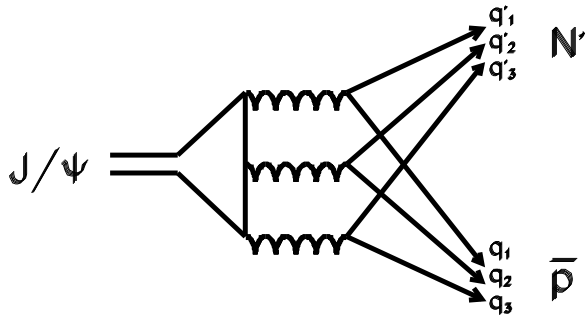}
\end{center}
\vspace*{-4.0cm}
\caption{Lowest-order diagram for $J/\Psi\to\bar pN'$ with $N'$ to be $p$
or $N^*$ }
\end{figure}

If $N^*(1440)$ is a pure three-quark baryon, $J/\Psi\to\bar pN^*(1440)$
should have the same Feynman diagram Fig.1 as for $J/\Psi\to\bar pp$.
The only difference for the two processes is their quark wave functions
and masses. Here we perform a calculation of the ratio between production
rates of two processes by assuming simple three-quark wave functions
for them. By comparing with experimental data, we can see whether the 
$N^*(1440)$ is produced more than the quark model prediction.

\section{Formalism}

For the basic amplitude corresponding to Fig.1, we have

\begin{eqnarray}
& & \langle q_i,s_i,q'_i,s'_i, i=1,2,3| \hat T |J/\Psi^{(\Lambda)}\rangle
\nonumber\\
& = & C_0\delta^4(P_\Psi-\sum_{i=1}^3q_i-\sum_{i=1}^3q'_i)
\cdot\epsilon_\Psi^{(\Lambda)\lambda}\cdot
{g_{\mu\lambda}g_{\nu\rho}+g_{\nu\lambda}g_{\mu\rho}
+g_{\rho\lambda}g_{\mu\nu}
\over (q_1+q'_1)^2(q_2+q'_2)^2(q_3+q'_3)^2}
\nonumber \\ & & \cdot
\bar u(q'_1,s'_1)\gamma^\mu v(q_1,s_1)\bar u(q'_2,s'_2)\gamma^\nu v(q_2,s_2)
\bar u(q'_3,s'_3)\gamma^\rho v(q_3,s_3)
\end{eqnarray}
where $\epsilon_\Psi^{(\Lambda)}$ is the polarization four-vector of
$J/\Psi$ with the helicity value $\Lambda$, $P_\Psi$ is the four-vector
momentum of $J/\Psi$, $q'_i,s'_i (q_i,s_i)$ 
are the four-vector momenta and spin z-projection of quarks 
(anti-quarks), respectively. We have put all color matrix elements, QCD
strong coupling constants, $J/\Psi$ decay constants,
etc., into a single overall constant $C_0$.

The relation between the $J/\Psi\to\bar pN'$  amplitude and the basic 
quark diagram amplitude Eq.(1) is
\begin{eqnarray}
{\cal M}^{(\Lambda)}_{s_z,s'_z}&\equiv & 
\langle\Psi_{\bar p}(q,s_z)\Psi_{N'}(q',s'_z)| \hat T
|J/\Psi^{(\Lambda)}\rangle
\nonumber\\
& = &  \sum_{s_i,s'_i}\int\prod_{i=1}^3\frac{d\vec q_i}{(2\pi)^32q_i^0}
\frac{d\vec q'_i}{(2\pi)^32q^{\prime 0}_i}  
\langle\Psi_{\bar p}(q,s_z)\Psi_{N'}(q',s'_z)|q_i,s_i,q'_i,s'_i,
i=1,2,3\rangle
\nonumber\\
& & \cdot \langle q_i,s_i,q'_i,s'_i, i=1,2,3| \hat T
|J/\Psi^{(\Lambda)}\rangle.
\end{eqnarray}
Here $\langle\Psi_{\bar p}(q,s_z)\Psi_{N'}(q',s'_z)|q_i,s_i,q'_i,s'_i,
i=1,2,3\rangle$
is the product of quark model wave functions of $\bar p$ and $N'$ in
momentum space, with constraints $\delta^4(q-q_1-q_2-q_3)\cdot
\delta^4(q'-q'_1-q'_2-q'_3)$.
The only difference between quark wave functions of the
proton and $N^*(1440)$ is their spatial parts, which we assume to be
simple harmonic-oscillator eigenfunctions in their center-of-mass
(CM) systems, {\sl i.e.},
\begin{eqnarray}
\Phi_{\bar p}(\vec k_\rho,\vec k_\lambda) & = & ({1\over\pi\alpha})^{3/2}
e^{-{1\over 2\alpha}(\vec k^2_\rho+\vec k^2_\lambda)} 
\quad for \quad proton; \\
\Phi_{N^*}(\vec k_\rho,\vec k_\lambda) & = &
\sqrt{3}({1\over\pi\alpha})^{3/2} 
[1-{1\over 3\alpha}(\vec k^2_\rho+\vec k^2_\lambda)]
e^{-{1\over 2\alpha}(\vec k^2_\rho+\vec k^2_\lambda)}
\quad for \quad N^*(1440)
\end{eqnarray}
where $\alpha=m\omega$ is the harmonic-oscillator parameter,
\begin{eqnarray}
\vec k_\rho &=& {1\over\sqrt{6}}(\vec k_1+\vec k_2 - 2\vec k_3) , \\
\vec k_\lambda &=& {1\over\sqrt{2}}(\vec k_1-\vec k_2), 
\end{eqnarray}
with $\vec k_1$, $\vec k_2$, and $\vec k_3$  the three quark momenta in
the CM system of their corresponding baryon, which are related to 
$\vec q_i$ or $\vec q'_i$ by a Lorentz transformation. 

In the $J/\Psi$ at rest system, the two baryon clusters are moving in
opposite directions with highly relativistic speeds, each becoming very
flat. Their spatial quark wave functions in this system are related to
their CM wave functions as follows\cite{Keister}:

\begin{equation}
\Psi(\vec q_\rho, \vec q_\lambda)=\left\vert
{\partial(\vec k_\rho, \vec k_\lambda)\over 
\partial(\vec q_\rho, \vec q_\lambda)}\right\vert^{1/2}  
\Phi(\vec k_\rho,\vec k_\lambda) 
\end{equation}
where
\begin{eqnarray}
\vec q_\rho &=& {1\over\sqrt{6}}(\vec q_1+\vec q_2 - 2\vec q_3) , \\
\vec q_\lambda &=& {1\over\sqrt{2}}(\vec q_1-\vec q_2) .
\end{eqnarray}

The spin and flavor wavefunctions of the proton and $N^*(1440)$  are the
same, {\sl i.e.},
\begin{equation}
\Psi^1_{SF}=\Psi^2_{SF}={1\over\sqrt{2}}(\chi^\rho\phi^\rho
+\chi^\lambda\phi^\lambda)
\end{equation}
where $\chi^\rho$ and $\chi^\lambda$ are the mixed-symmetry pair
spin-${1\over 2}$ wavefunctions.  For example, we have
\begin{eqnarray}
\chi^\rho_{{1\over 2},{1\over 2}} &=& -{1\over\sqrt{6}}
\{|\uparrow\downarrow\uparrow>+|\downarrow\uparrow\uparrow>
-2|\uparrow\uparrow\downarrow>\}, \\
\chi^\lambda_{{1\over 2},{1\over 2}} &=& {1\over\sqrt{2}}
\{|\uparrow\downarrow\uparrow>-|\downarrow\uparrow\uparrow>\}
\end{eqnarray}
for the case of the total spin ${1\over 2}$ and its projection 
${1\over 2}$. The flavor wavefunctions $\phi^\rho$ and $\phi^\lambda$
are exactly analogous to that of the spin wavefunctions but in
isospin space of $u$-$d$ quarks. 

We perform the calculation in the $J/\Psi$ rest system.
For $J/\Psi$ produced in $e^+e^-$ annihilation, its helicity is
limited to be $\Lambda=\pm 1$.
The components in Eq.(1) can be expressed more explicitly as
\begin{eqnarray}
& & \epsilon^{(\pm)}_\Psi = (0; \mp{1\over\sqrt{2}},-{i\over\sqrt{2}},0) ,\\
& & \bar u(q'_i,s'_i)\gamma^0v(q_i,s_i) = 0, \\
& & \bar u(q'_i,s'_i)\vec\gamma v(q_i,s_i) = {E_q+m_q\over 2m_q}
\langle s'_i\left\vert(1+{|\vec q_i|^2\over (E_q+m_q)^2})\vec\sigma
-{2(\vec\sigma\cdot\vec q_i)\vec q_i\over (E_q+m_q)^2}\right\vert
s_i\rangle
\end{eqnarray} 
where $E_q$ and $m_q$ are the energy and mass of the quark.

In Eq.(2), the integration over $\prod_{i=1}^3d\vec q_id\vec q'_i$ with two
$\delta^4$ functions can be reduced to a ten-dimension integration which
we carry out numerically with the adaptive multidimensional Monte-Carlo
integration program RIWIAD of CERN Program Library. From these amplitudes
${\cal M}^{(\Lambda)}_{s_z,s'_z}$, we can get the decay cross section for
$J/\Psi^{(\Lambda)}\to\bar pN'$ as
\begin{equation}
d\Gamma(J/\Psi^{(\Lambda)}\to\bar pN')={1\over 32\pi^2}
\{|{\cal M}^{(\Lambda)}_{{1\over 2},{1\over 2}}|^2+
|{\cal M}^{(\Lambda)}_{{1\over 2},-{1\over 2}}|^2+
|{\cal M}^{(\Lambda)}_{-{1\over 2},{1\over 2}}|^2+
|{\cal M}^{(\Lambda)}_{-{1\over 2},-{1\over 2}}|^2\}{|\vec q|\over
M_\Psi^2}d\Omega
\end{equation}
with $\Omega$ as the solid angle of $\vec q$.

The calculation of $J/\Psi\to \bar N^*N^*$ is similar, just replacing
quark radial wavefunction of the anti-proton by that of the $\bar
N^*(1440)$. With formulas above, the calculation of the decay cross
sections is straightforward though tedious.

\section{Numerical results and discussion}

In our quark model calculation, there are three parameters, {\sl i.e.},
the constituent quark mass $m_q$, the harmonic-oscillator parameter
$\alpha$ and an overall normalization factor $C_0$. The relation between
$\alpha$ and the nucleon radius $r_0$ is $\alpha=1/r_0^2$.
In most quark model 
calculations\cite{Isgur1,Keister,Foster,Capstick,Dong}, the quark mass
$m_q$ has been chosen in the range of $220\sim 340$ MeV, and $\alpha$ in
the range of $0.06\sim 0.22 GeV^2$ which corresponds to the nucleon radius 
in the range of $0.42\sim 0.8$ fm. In the following, we limit our
parameters in these ranges.

\begin{figure}[htbp]
\begin{center}\hspace*{-0.cm}
\epsfysize=10cm
\epsffile{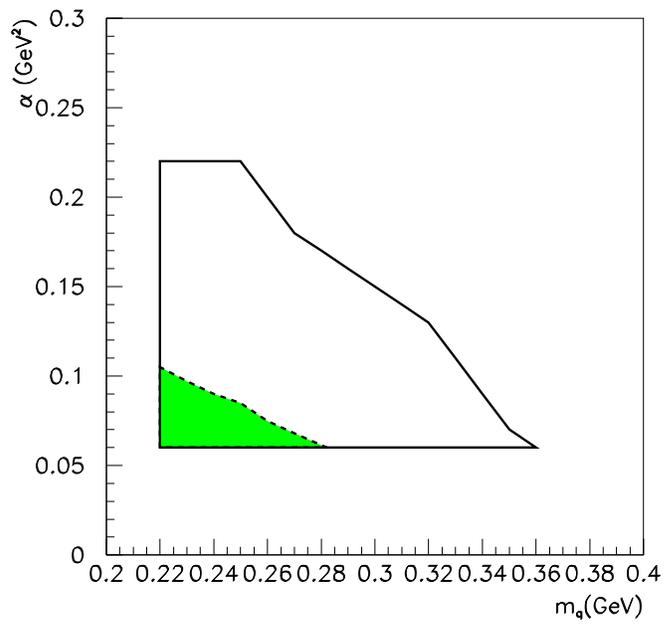}
\caption{The constrained area for parameters $(\alpha, m_q)$ 
from experimental data $\alpha_p=0.62\pm 0.11$\cite{DM2}. 
The shaded area is the result with the Lorentz contraction effect;
the area surrounded by the solid line is the result ignoring the 
Lorentz contraction effect. }
\label{fig2}
\end{center}
\end{figure}

The $J/\Psi$ decay cross sections for $\bar pp$, $\bar pN^*$ and 
$\bar N^*N^{*+}$ can be expressed as 
\begin{eqnarray}
{d\Gamma(J/\Psi^{(\pm)}\to\bar pp)\over d\Omega} & = & 
N_{\bar pp}(1+\alpha_p cos^2\theta) ,\\
{d\Gamma(J/\Psi^{(\pm)}\to\bar pN^*)\over d\Omega} & = & 
R_* N_{\bar pp}(1+\alpha_* cos^2\theta) , \\
{d\Gamma(J/\Psi^{(\pm)}\to\bar N^*N^{*+})\over d\Omega} & = &
R_{**}N_{\bar pp}(1+\alpha_{**} cos^2\theta). 
\end{eqnarray}
Here $N_{\bar pp}$ is a constant direct related to the experimental
branching ratio of $J/\Psi\to\bar pp$ and can be used to fix the overall
normalization constant $C_0$.
The experimental value for $\alpha_p$ is ($0.62\pm 0.11$)\cite{DM2} and
can be used to put further limit on the range of parameters $\alpha$ and
$m_q$. The shaded area in Fig.\ref{fig2} shows the range allowed by one
standard deviation of the experimental $\alpha_p$ value. 

In order to investigate the importance of the Lorentz contraction effect,
we have also performed the calculation by ignoring this effect, 
{\sl i.e.},
assuming $\vec k_\rho=\vec q_\rho$ and $\vec k_\lambda=\vec q_\lambda$.
The resulted ($\alpha$, $m_q$) area allowed by one
standard deviation of the experimental $\alpha_p$ value is shown
in Fig.\ref{fig2} by the area surrounded by the solid line.
One can see that the Lorentz contraction effect is very large and
cannot be ignored.

With ($\alpha$, $m_q$) values in the shaded area of Fig.\ref{fig2}, 
our quark model calculation predicts
$\alpha_*=0.36\pm 0.08$, $\alpha_{**}=0.08\pm 0.05$,  
$R_*=2.1\sim 4.8$ and $R_{**}=2.0\sim 24.0$. 
Mixings between the ground state and the radially excited
states\cite{Capstick} will not change our result much due to the relative
negative sign of mixings for the proton and $N^*(1440)$.

There are no experimental data on $\bar pN^*$ and $\bar N^*N^*$ channels
yet. However from both BESI\cite{BES} and MARKII\cite{Mark2} experiments,
there is a clear peak around 1.5 GeV in the $\pi N$ invariant mass in 
$J/\psi\to\bar pN\pi$ processes, although no partial wave analyses were
performed. Very recently BESII has finished data-taking for 50 million
more $J/\psi$ events, which is about two order of manitude more statistics
than MARKII data and one order of magnitude more statistics than BESI
data. With such statistics, partial wave analyses of relevant channels
are possible.
New experimental results on $J/\Psi\to\bar pp$, $J/\Psi\to\bar pN^*$
and $J/\Psi\to\bar N^*N^*$ will help us to narrow down the quark
model $(\alpha, m_q)$ parameters and study the nature of $N^*$. If the
$J/\Psi\to\bar pN^*$ production rate
is significantly larger than our quark model prediction, it may indicate
that the $N^*$ is a hybrid\cite{Page}; if $J/\Psi\to\bar pN^*$ production
rate is significantly smaller than our prediction, then it may indicate
that the $N^*$ contains a large component of $\pi N$ in its internal
structure\cite{Dong}. For a more quantitative statement, concrete
theoretical calculations for hybrid and molecule baryon production are
needed.

\medskip
{\bf Acknowledgement:} 
We thank D.V.Bugg, Y.B.Dong, L.Kisslinger, H.B.Li, Y.W.Yu and Z.Y.Zhang 
for useful discussions and comments. This work is partly supported by
National Science Foundation of China under contract Nos. 19991487 and
19905011.

\end{document}